\documentclass[aps,prd,twocolumn,showpacs,superscriptaddress,letterpaper,amsmath,amssymb]{revtex4} 
\usepackage{graphicx}
\usepackage{dcolumn}
\usepackage{bm}
\usepackage{float}

\RequirePackage{xspace}
\usepackage{relsize}

\def\Kbar {\kern 0.2em\overline{\kern -0.2em K}{}\xspace}

\def\Bbar {\kern 0.18em\overline{\kern -0.18em B}{}\xspace}

\def\Qbar {\kern 0.08em\overline{\kern -0.08em Q}{}\xspace}

\newcommand{\mev}{\ensuremath{\mathrm{\,Me\kern -0.1em V}}\xspace}
\newcommand{\mevc}{\ensuremath{{\mathrm{\,Me\kern -0.1em V\!/}c}}\xspace}
\newcommand{\mevcc}{\ensuremath{{\mathrm{\,Me\kern -0.1em V\!/}c^2}}\xspace}
\newcommand{\gev}{\ensuremath{\mathrm{\,Ge\kern -0.1em V}}\xspace}
\newcommand{\gevc}{\ensuremath{{\mathrm{\,Ge\kern -0.1em V\!/}c}}\xspace}
\newcommand{\gevcnospace}{\ensuremath{{\mathrm{\,Ge\kern -0.1em V\!/}c}}}
\newcommand{\gevcc}{\ensuremath{{\mathrm{\,Ge\kern -0.1em V\!/}c^2}}\xspace}

\def\missET {\mbox{\ensuremath{\not\!\! E_T}}}
\begin{document}
\bibliographystyle{apsrev}

\title{Search for new phenomena in events with two $Z$ bosons and
 missing transverse momentum in $p\bar{p}$ collisions at $\sqrt{s}=1.96$~TeV}
\affiliation{Institute of Physics, Academia Sinica, Taipei, Taiwan 11529, Republic of China}
\affiliation{Argonne National Laboratory, Argonne, Illinois 60439, USA}
\affiliation{University of Athens, 157 71 Athens, Greece}
\affiliation{Institut de Fisica d'Altes Energies, ICREA, Universitat Autonoma de Barcelona, E-08193, Bellaterra (Barcelona), Spain}
\affiliation{Baylor University, Waco, Texas 76798, USA}
\affiliation{Istituto Nazionale di Fisica Nucleare Bologna, $^{ee}$University of Bologna, I-40127 Bologna, Italy}
\affiliation{University of California, Davis, Davis, California 95616, USA}
\affiliation{University of California, Irvine, California 92627, USA}
\affiliation{University of California, Los Angeles, Los Angeles, California 90024, USA}
\affiliation{Instituto de Fisica de Cantabria, CSIC-University of Cantabria, 39005 Santander, Spain}
\affiliation{Carnegie Mellon University, Pittsburgh, Pennsylvania 15213, USA}
\affiliation{Enrico Fermi Institute, University of Chicago, Chicago, Illinois 60637, USA}
\affiliation{Comenius University, 842 48 Bratislava, Slovakia; Institute of Experimental Physics, 040 01 Kosice, Slovakia}
\affiliation{Joint Institute for Nuclear Research, RU-141980 Dubna, Russia}
\affiliation{Duke University, Durham, North Carolina 27708, USA}
\affiliation{Fermi National Accelerator Laboratory, Batavia, Illinois 60510, USA}
\affiliation{University of Florida, Gainesville, Florida 32611, USA}
\affiliation{Laboratori Nazionali di Frascati, Istituto Nazionale di Fisica Nucleare, I-00044 Frascati, Italy}
\affiliation{University of Geneva, CH-1211 Geneva 4, Switzerland}
\affiliation{Glasgow University, Glasgow G12 8QQ, United Kingdom}
\affiliation{Harvard University, Cambridge, Massachusetts 02138, USA}
\affiliation{Division of High Energy Physics, Department of Physics, University of Helsinki and Helsinki Institute of Physics, FIN-00014, Helsinki, Finland}
\affiliation{University of Illinois, Urbana, Illinois 61801, USA}
\affiliation{The Johns Hopkins University, Baltimore, Maryland 21218, USA}
\affiliation{Institut f\"{u}r Experimentelle Kernphysik, Karlsruhe Institute of Technology, D-76131 Karlsruhe, Germany}
\affiliation{Center for High Energy Physics: Kyungpook National University, Daegu 702-701, Korea; Seoul National University, Seoul 151-742, Korea; Sungkyunkwan University, Suwon 440-746, Korea; Korea Institute of Science and Technology Information, Daejeon 305-806, Korea; Chonnam National University, Gwangju 500-757, Korea; Chonbuk National University, Jeonju 561-756, Korea}
\affiliation{Ernest Orlando Lawrence Berkeley National Laboratory, Berkeley, California 94720, USA}
\affiliation{University of Liverpool, Liverpool L69 7ZE, United Kingdom}
\affiliation{University College London, London WC1E 6BT, United Kingdom}
\affiliation{Centro de Investigaciones Energeticas Medioambientales y Tecnologicas, E-28040 Madrid, Spain}
\affiliation{Massachusetts Institute of Technology, Cambridge, Massachusetts 02139, USA}
\affiliation{Institute of Particle Physics: McGill University, Montr\'{e}al, Qu\'{e}bec, Canada H3A~2T8; Simon Fraser University, Burnaby, British Columbia, Canada V5A~1S6; University of Toronto, Toronto, Ontario, Canada M5S~1A7; and TRIUMF, Vancouver, British Columbia, Canada V6T~2A3}
\affiliation{University of Michigan, Ann Arbor, Michigan 48109, USA}
\affiliation{Michigan State University, East Lansing, Michigan 48824, USA}
\affiliation{Institution for Theoretical and Experimental Physics, ITEP, Moscow 117259, Russia}
\affiliation{University of New Mexico, Albuquerque, New Mexico 87131, USA}
\affiliation{The Ohio State University, Columbus, Ohio 43210, USA}
\affiliation{Okayama University, Okayama 700-8530, Japan}
\affiliation{Osaka City University, Osaka 588, Japan}
\affiliation{University of Oxford, Oxford OX1 3RH, United Kingdom}
\affiliation{Istituto Nazionale di Fisica Nucleare, Sezione di Padova-Trento, $^{ff}$University of Padova, I-35131 Padova, Italy}
\affiliation{University of Pennsylvania, Philadelphia, Pennsylvania 19104, USA}
\affiliation{Istituto Nazionale di Fisica Nucleare Pisa, $^{gg}$University of Pisa, $^{hh}$University of Siena and $^{ii}$Scuola Normale Superiore, I-56127 Pisa, Italy}
\affiliation{University of Pittsburgh, Pittsburgh, Pennsylvania 15260, USA}
\affiliation{Purdue University, West Lafayette, Indiana 47907, USA}
\affiliation{University of Rochester, Rochester, New York 14627, USA}
\affiliation{The Rockefeller University, New York, New York 10065, USA}
\affiliation{Istituto Nazionale di Fisica Nucleare, Sezione di Roma 1, $^{jj}$Sapienza Universit\`{a} di Roma, I-00185 Roma, Italy}
\affiliation{Rutgers University, Piscataway, New Jersey 08855, USA}
\affiliation{Texas A\&M University, College Station, Texas 77843, USA}
\affiliation{Istituto Nazionale di Fisica Nucleare Trieste/Udine, I-34100 Trieste, $^{kk}$University of Udine, I-33100 Udine, Italy}
\affiliation{University of Tsukuba, Tsukuba, Ibaraki 305, Japan}
\affiliation{Tufts University, Medford, Massachusetts 02155, USA}
\affiliation{University of Virginia, Charlottesville, Virginia 22906, USA}
\affiliation{Waseda University, Tokyo 169, Japan}
\affiliation{Wayne State University, Detroit, Michigan 48201, USA}
\affiliation{University of Wisconsin, Madison, Wisconsin 53706, USA}
\affiliation{Yale University, New Haven, Connecticut 06520, USA}

\author{T.~Aaltonen}
\affiliation{Division of High Energy Physics, Department of Physics, University of Helsinki and Helsinki Institute of Physics, FIN-00014, Helsinki, Finland}
\author{B.~\'{A}lvarez~Gonz\'{a}lez$^z$}
\affiliation{Instituto de Fisica de Cantabria, CSIC-University of Cantabria, 39005 Santander, Spain}
\author{S.~Amerio}
\affiliation{Istituto Nazionale di Fisica Nucleare, Sezione di Padova-Trento, $^{ff}$University of Padova, I-35131 Padova, Italy}
\author{D.~Amidei}
\affiliation{University of Michigan, Ann Arbor, Michigan 48109, USA}
\author{A.~Anastassov$^x$}
\affiliation{Fermi National Accelerator Laboratory, Batavia, Illinois 60510, USA}
\author{A.~Annovi}
\affiliation{Laboratori Nazionali di Frascati, Istituto Nazionale di Fisica Nucleare, I-00044 Frascati, Italy}
\author{J.~Antos}
\affiliation{Comenius University, 842 48 Bratislava, Slovakia; Institute of Experimental Physics, 040 01 Kosice, Slovakia}
\author{G.~Apollinari}
\affiliation{Fermi National Accelerator Laboratory, Batavia, Illinois 60510, USA}
\author{J.A.~Appel}
\affiliation{Fermi National Accelerator Laboratory, Batavia, Illinois 60510, USA}
\author{T.~Arisawa}
\affiliation{Waseda University, Tokyo 169, Japan}
% ...

\collaboration{CDF Collaboration\footnote{With visitors from
$^a$Istituto Nazionale di Fisica Nucleare, Sezione di Cagliari, 09042 Monserrato (Cagliari), Italy,
$^b$University of CA Irvine, Irvine, CA 92697, USA,
$^c$University of CA Santa Barbara, Santa Barbara, CA 93106, USA,
$^d$University of CA Santa Cruz, Santa Cruz, CA 95064, USA,
$^e$Institute of Physics, Academy of Sciences of the Czech Republic, Czech Republic,
$^f$CERN, CH-1211 Geneva, Switzerland,
$^g$Cornell University, Ithaca, NY 14853, USA,
$^h$University of Cyprus, Nicosia CY-1678, Cyprus,
$^i$Office of Science, U.S. Department of Energy, Washington, DC 20585, USA,
$^j$University College Dublin, Dublin 4, Ireland,
$^k$ETH, 8092 Zurich, Switzerland,
$^l$University of Fukui, Fukui City, Fukui Prefecture, Japan 910-0017,
$^m$Universidad Iberoamericana, Mexico D.F., Mexico,
$^n$University of Iowa, Iowa City, IA 52242, USA,
$^o$Kinki University, Higashi-Osaka City, Japan 577-8502,
$^p$Kansas State University, Manhattan, KS 66506, USA,
$^q$Korea University, Seoul, 136-713, Korea,
$^r$University of Manchester, Manchester M13 9PL, United Kingdom,
$^s$Queen Mary, University of London, London, E1 4NS, United Kingdom,
$^t$University of Melbourne, Victoria 3010, Australia,
$^u$Muons, Inc., Batavia, IL 60510, USA,
$^v$Nagasaki Institute of Applied Science, Nagasaki, Japan,
$^w$National Research Nuclear University, Moscow, Russia,
$^x$Northwestern University, Evanston, IL 60208, USA,
$^y$University of Notre Dame, Notre Dame, IN 46556, USA,
$^z$Universidad de Oviedo, E-33007 Oviedo, Spain,
$^{aa}$CNRS-IN2P3, Paris, F-75205 France,
$^{bb}$Texas Tech University, Lubbock, TX 79609, USA,
$^{cc}$Universidad Tecnica Federico Santa Maria, 110v Valparaiso, Chile,
$^{dd}$Yarmouk University, Irbid 211-63, Jordan,
}}
\noaffiliation

\begin{abstract}
We present a search for new phenomena in events with two reconstructed
$Z$ bosons and large missing transverse momentum, sensitive to
processes $p\bar{p}\rightarrow X_2X_2 \rightarrow Z Z X_1 X_1$, where $X_2$ is
an unstable particle decaying as $X_2\rightarrow ZX_1$ and $X_1$ is undetected. The particles $X_1$ and $X_2$
may be, among other possibilities, fourth generation neutrinos or supersymmetric particles. We study the final state in which one $Z$ boson decays to
two charged leptons and the second decays hadronically. In data corresponding to an integrated luminosity of 4.2
fb$^{-1}$ from proton-antiproton collisions recorded by the
CDF II detector at the Tevatron, with center-of-mass energy of 1.96 TeV,
we find agreement between data and standard-model backgrounds. We
calculate 95\% confidence level upper limits on the cross section of the process
$p\bar{p}\rightarrow X_2X_2 \rightarrow Z Z X_1 X_1$ ranging from 50 fb to 1
pb, depending on the masses of $X_1$ and $X_2$.
\end{abstract}


\maketitle

\date{\today}

A natural extension to the standard model of particle physics is a fourth generation of quarks and leptons. The inclusion of a fourth generation provides a source of $CP$ violation in $B_s$ decays and can accommodate a heavy Higgs boson~\cite{fourthgen,fourthgenconstraints}. Searches for fourth generation quarks at the Fermilab Tevatron have constrained the mass of up-type quarks ($u_4$), that decay as $u_{4} \rightarrow Wq$, where $q$ is a generic down-type quark, to be $m_{u_{4}} >$ 340 GeV$/c^2$ at 95\% confidence level (CL)~\cite{cdftprime}, while limits on the mass of down-type quarks ($d_{4}$) decaying via $d_{4} \rightarrow Wt$ are $m_{d4}>$ 372 GeV$/c^2$ at 95\% CL~\cite{cdfbprime}.

Following the trend of mass hierarchy in the standard model, the least massive and therefore most accessible particle of this fourth generation may be the neutrino. Such a neutrino need not be solely a Dirac or 
Majorana state, but may be a mixture of the two~\cite{nureana}. 
This leads to two mass eigenstates $N_1$ and $N_2$, where $N_{2}$ is the unstable heavy eigenstate and $N_{1}$ is the stable and least massive eigenstate of the fourth generation neutrinos. 
These particles would partially evade the neutrino mass constraints from $Z$ width studies at LEP~\cite{zwidthlep}.

The dominant production mechanism of $N_1$ would be via a Drell-Yan
process, $p\bar{p}\rightarrow Z/\gamma^* \rightarrow N_{2}N_{2}
\rightarrow N_{1}ZN_{1}Z$, giving a final state of two $Z$ bosons and
large missing transverse momentum. This signature is
shared by several other interesting new physics processes, most
notably supersymmetric production, $\chi^0_2\chi^0_2 \rightarrow
Z\chi^0_1Z\chi^0_1$, where $\chi^0_1$ and $\chi^0_2$ are neutralinos. We consider the mode in which one $Z$ decays
hadronically and the other decays leptonically, giving a detector
signature of two charged leptons, two jets and large missing
transverse momentum. For this search we use $p\bar{p}$ collisions at $\sqrt{s}=1.96$~TeV
 corresponding to 4.2 fb$^{-1}$ of integrated luminosity collected by the CDF II detector.

Events were recorded by  CDF II~\cite{cdf,coordinates}, a general-purpose detector designed to study collisions at the Fermilab Tevatron $p\overline{p}$ collider. The CDF II detector is composed of a charged-particle tracking system immersed in a 1.4 T magnetic field consisting of a silicon microstrip tracker and a drift chamber. Electromagnetic and hadronic calorimeters surround the tracking system and measure particle energies. Drift chambers located outside the calorimeters detect muons. 

The data acquisition system is triggered by $e$ or $\mu$ candidates with transverse momentum $p_T$, greater than $18$ GeV$/c$. We retain electron and muon candidates with
pseudorapidity~\cite{coordinates} $|\eta| < 1.1$, $p_T \ge 20$
GeV$/c$ and that satisfy the standard CDF identification requirements
~\cite{leptonid}. For muons, the track fit $\chi^2$ per degree of freedom is used to
reject poorly fit tracks likely resulting from charged pion and kaon decays in flight. Electrons from photon
conversions are suppressed by rejecting electron candidates with a nearly collinear
intersecting reconstructed track. Jets are reconstructed in the
calorimeter using the {\sc jetclu }~\cite{jetclu} algorithm with a
clustering radius of 0.4 in azimuth-pseudorapidity space. Measured jet energies are corrected to account
for $\eta$-dependent variations in detector response, calorimeter
coverage, and the expected contribution from additional $p\bar{p}$
interactions in the same event~\cite{jes}. Jets are selected if they
have $p_T \ge 15$ GeV$/c$ and $|\eta| < 2.4$. Missing transverse
energy~\cite{metdef}, $\missET$, is reconstructed using calorimeter
and muon information including the corrections described above.

To isolate the $ZZ$ signature, we require two opposite-charge,
same-flavor lepton candidates ($e$ or $\mu$) with $p_T >$ 20 GeV$/c$ for which
the lepton-pair invariant mass is consistent with decay from a $Z$ boson:
$m_{\ell\ell} \in [76,106]$ GeV$/c^2$. Additionally, we require at least two jets, each with
$p_T>15$ GeV$/c$ and\ $|\eta|<2.4$, and without identified
secondary vertices resulting from $b$-hadron decay~\cite{secvtx}. The
$ZZ+\missET$ signature has the further requirement of large \missET,
varying with hypothetical $N_1$ and $N_2$ masses, as shown in Table~\ref{tab:limits}.

\begin{figure}[!htpb]
\begin{center}
\includegraphics[width=1.0\linewidth]{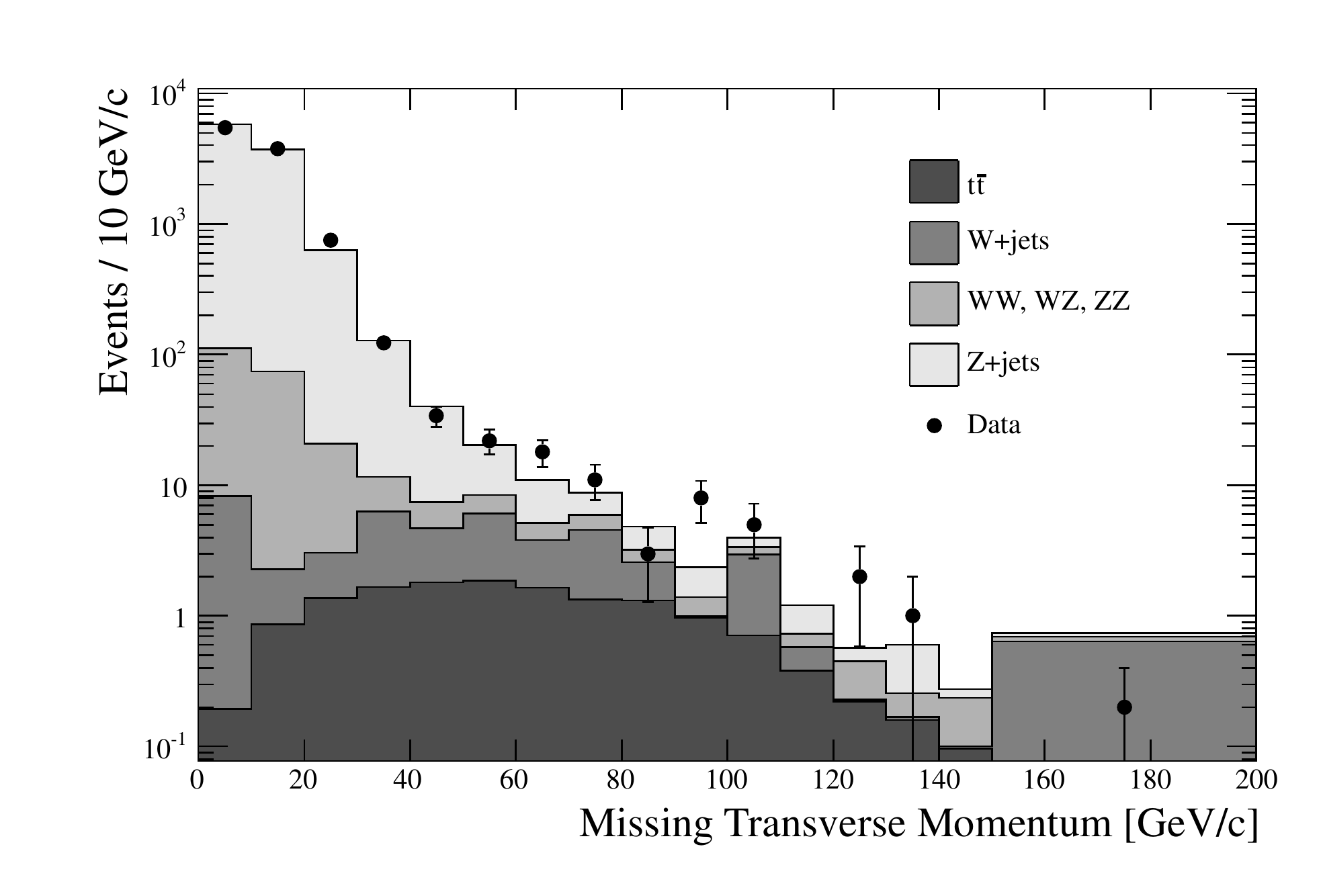}
\caption{ Distribution of missing transverse momentum in events with the
 $ZZ$ signature, for expected backgrounds and observed data.}
\label{fig:met}
\end{center}
\end{figure}

The dominant background in the resulting sample is production of a $Z$
boson in association with two jets from initial state radiation. We
model this background using {\sc alpgen}~\cite{alpgen} to describe the
hard process and {\sc pythia}~\cite{pythia} for the showering and
hadronization. This background is strongly suppressed in events with large missing transverse momentum, as shown in
Figure~\ref{fig:met} and Table~\ref{tab:sigbg},
and is distinguished from the signal by the lack of a resonance in the dijet mass, $m_{jj}$.

The second largest expected background is due to $W$ boson production in association with three jets from initial state radiation, where one jet is wrongly reconstructed as a lepton. We model this using an independent sample of events containing jets likely to mimic leptons, following Ref.~\cite{zhllbb}. Additional backgrounds result from standard-model production of two gauge bosons, including $ZZ$, $WW$, and $WZ$, as well as $t\bar{t}\rightarrow WbWb$, which are all modeled using {\sc pythia}.

\begin{table}[!htpb]
\caption{Expected number of events for each source of background to
 the $ZZ\rightarrow\ell^+\ell^{-}jj$ and $ZZ+X_1X_1\rightarrow\ell^+\ell^{-}jj+\missET$ signatures, as well as the observed event yield
 in data with 4.2 fb$^{-1}$ of integrated luminosity. The threshold in
 $\missET$ is optimized as a function of the $N_1,N_2$ masses; one
 example $(N_{1}=125$ GeV$/c^{2}, N_{2}=225$ GeV$/c^{2})$ is shown here. 
 Uncertainties shown include both systematic and statistical uncertainty added in quadrature. }
\begin{center}
\begin{ruledtabular}
\begin{tabular}{lcc} 
 & & $\ell^+\ell^{-}jj$ and \\
Process& $\ell^+\ell^{-}jj$ &\ \ \ \ \ $\missET>36$ GeV \\
\hline 
 $WW$ & 4.4 $\pm$ 1.3 & 2.7 $\pm$ 0.8 \\ 
 $t\bar{t}$ & 14.8 $\pm$ 3.0 & 11.6 $\pm$ 2.3 \\ 
 $W$+jets & 36.1 $\pm$ 16.7 & 21.7 $\pm$ 12.6 \\ 
 $ZZ$ & 99.4 $\pm$ 20.5 & 4.2 $\pm$ 0.9 \\ 
 $WZ$ & 105.6 $\pm$ 22.1 & 5.2 $\pm$ 1.1 \\ 
 $Z$+jets & 10171 $\pm$ 4422 & 94.6 $\pm$ 38.5 \\ 
\hline 
Total & 10432 $\pm$ 4485 & 140.0 $\pm$ 40.6 \\ 
Data & 10199 & 152 \\ 
\end{tabular} 
\label{tab:sigbg}
\end{ruledtabular}
\end{center} 
\end{table}

\begin{figure}[!htpb]
\begin{center}
\includegraphics[width=1.0\linewidth]{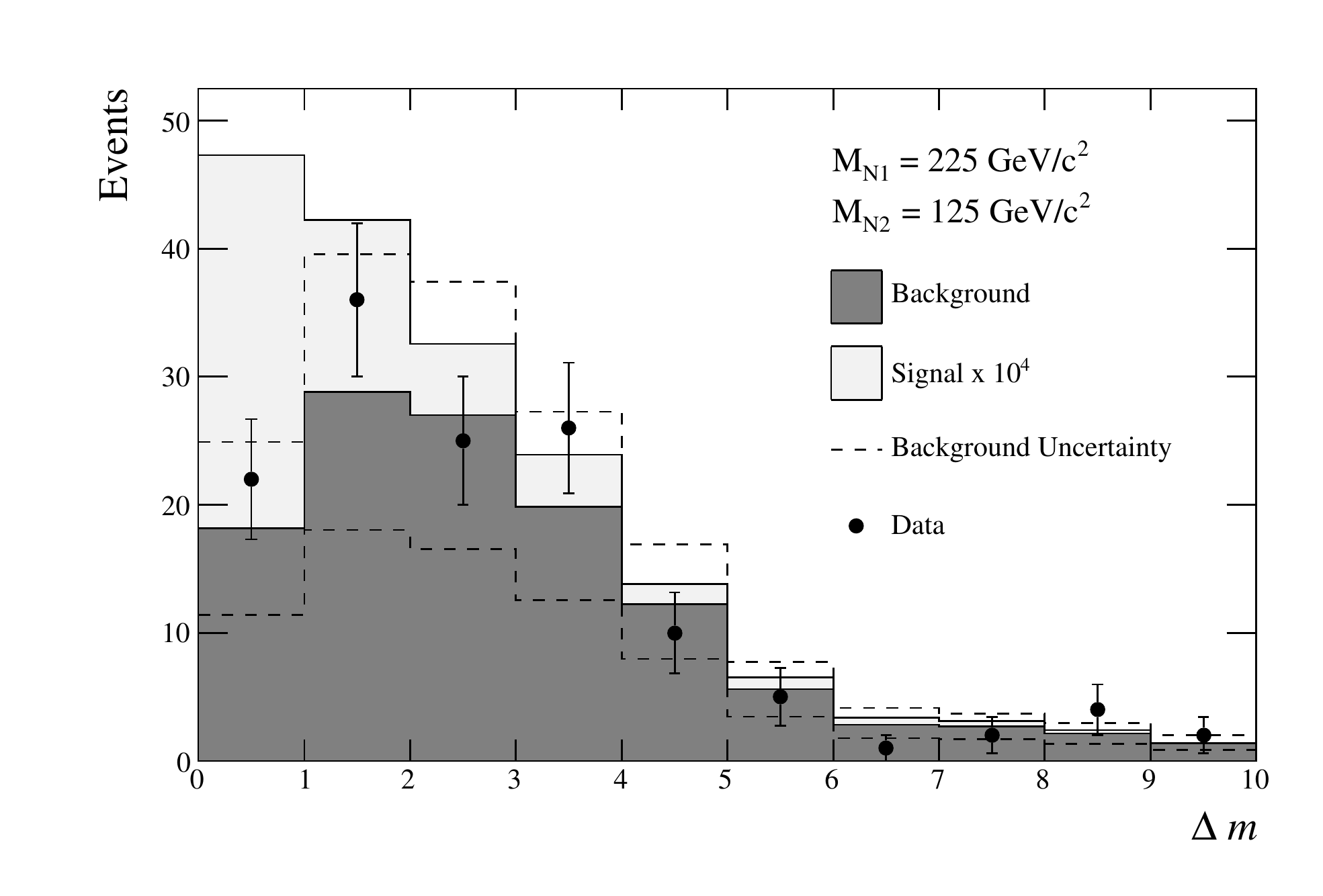}
\caption{ Distribution of the variable $\Delta m$, defined in the text,
 for expected background, observed data and an example signal (scaled by $10^{4}$) in data
 with 4.2 fb$^{-1}$ of integrated luminosity. This example uses
 a missing transverse momentum threshold of $\missET > 36$ GeV,
 optimized for this $(M_{N1},M_{N2})$ mass point; see
 Table~\ref{tab:limits}. Background uncertainties are statistical
 and systematic added in quadrature.}
\label{fig:dm}
\end{center}
\end{figure}

To isolate the double-resonance nature of the $ZZ+\missET$ signature, we calculate the distance from the $Z$ boson reconstructed mass in the $m_{\ell\ell}-m_{jj}$ mass plane, accounting for the relative difference in the resolutions between the leptons and jets as well as the observed bias in reconstructed $m_{jj}$, using the variable
\begin{equation}
\Delta m = \sqrt{\Big(\frac{m_{\ell\ell}-m_{Z\rightarrow\ell\ell}}{g_{\ell\ell}}\Big)^2+\Big(\frac{m_{jj}-m_{Z\rightarrow jj}}{g_{jj}}\Big)^2, }
\label{eq:deltamass}
\end{equation} 
\noindent 
where $m_{\ell\ell} (m_{jj})$ is the reconstructed lepton (jet) pair mass, compared
to the reference $m_{Z\rightarrow\ell\ell} = 91.6$ GeV$/c^2$
($m_{Z\rightarrow jj} = 85.3$ GeV$/c^2$) found in simulated events. To
account for the superior lepton resolution, the dilepton and dijet
mass differences are scaled by factors related to the resolutions:
$g_{\ell\ell} = 10$ GeV$/c^2$, $g_{jj} = 15$ GeV$/c^2$.  The uncertainties
of these reference values are small, and may be neglected. The distribution of $\Delta m$ for data and simulated background and signal is shown in Figure~\ref{fig:dm}.

\begin{figure}[!htpb]
\begin{center}
\includegraphics[width=0.95\linewidth]{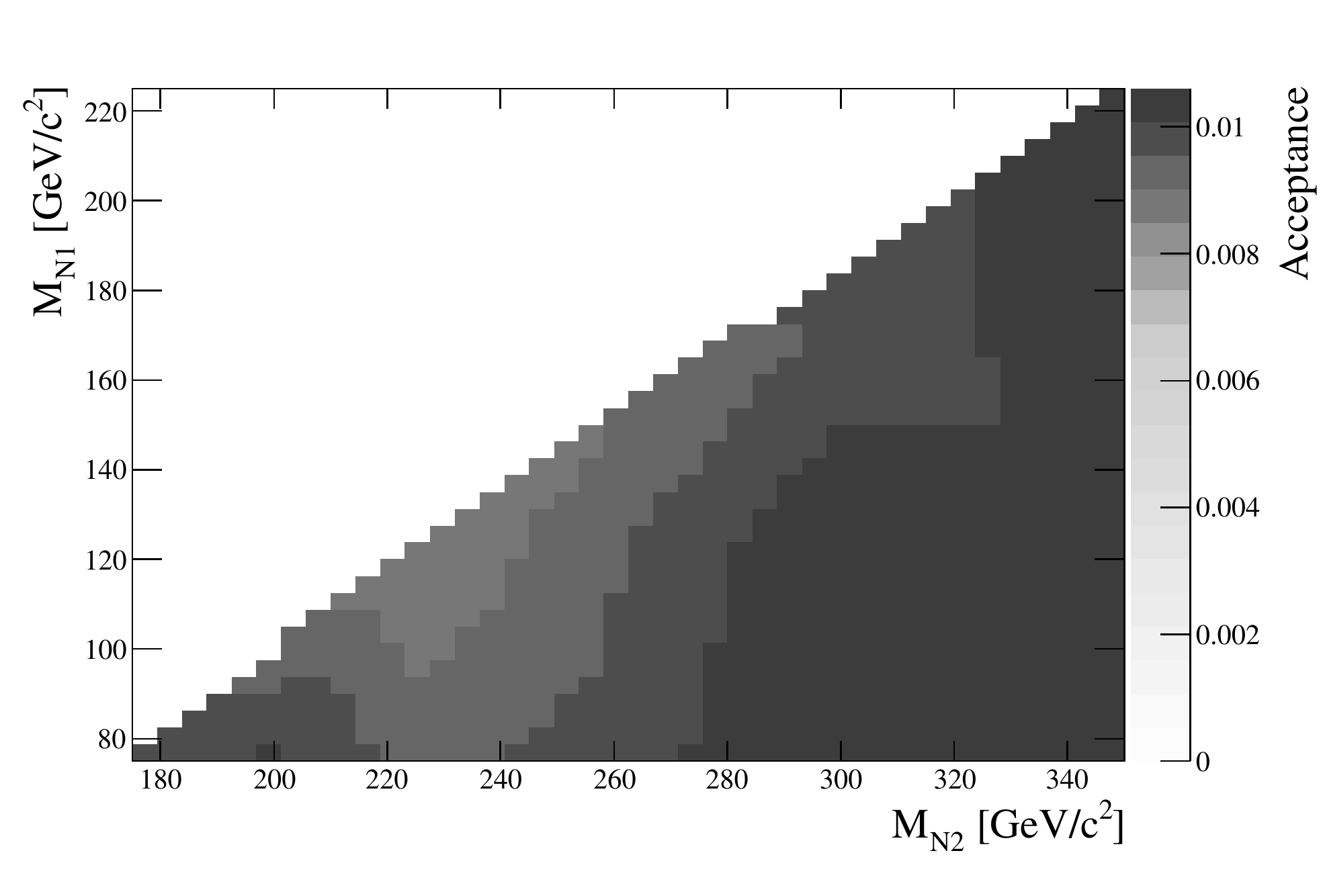}
\caption{ Acceptance of the $ZZ+\missET$ signature,
 including BR($ZZ\rightarrow \ell\ell qq$), as a function of the
 masses of the fourth generation neutrinos, $N_1$ and
 $N_2$. The threshold in $\missET$ is optimized at each point on a grid in this
 plane. Linear interpolation is performed between the grid points. The apparent structure in the plot results from statistical fluctuation.  \vspace{1mm}}
\label{fig:acc}
\end{center}
\end{figure}

We model the production of the $N_2$ signal and its subsequent decay
into $N_1$ over a grid of masses in the ($M_{N1},M_{N2}$) plane using
{\sc madgraph}~\cite{madgraph} with the {\sc cteq5l}~\cite{cteq} parton
distribution fuctions; {\sc pythia}~\cite{pythia} is
used for the showering and
hadronization. To suppress the large backgrounds expected from
standard-model sources we require large $\missET$; as the expected
magnitude of missing transverse momentum depends strongly on $M_{N1}$
and $M_{N2}$, we vary the selection threshold of $\missET$ to optimize
for sensitivity at each ($M_{N1},M_{N2}$) pair considered, as seen in
Table~\ref{tab:limits}. The acceptance for each mass point can be seen in Figure~\ref{fig:acc}. For each point in the mass grid, we form template histograms as a
function of $\Delta m$ for the expected signal and
background, as displayed in Figure~\ref{fig:dm}. 

In addition to the templates formed for the nominal expectation, we form alternate templates that incorporate the effects of systematic uncertainties under $\pm1\sigma$ variation. Fitting to these templates using
the maximum likelihood method, we extract the best-fit signal cross
section, $\sigma_{N2}$. Systematic uncertainties affecting the shapes
of templates, including uncertainty in the jet energy
scale~\cite{jes}, QCD radiation, PDFs, $Q^2$ (square of momentum
transfer in the interaction) and uncertainty in lepton energy
resolution, are accounted for as nuisance parameters in our
likelihood. The dominant source of systematic uncertainty in this
analysis is uncertainty in the jet energy scale (40\%), which can
significantly modify the number of jets in background processes that
pass the $p_T$ threshold, the location of the $m_{jj}$
resonance in the signal process, and the measured $\missET$ in an
event. The second largest systematic uncertainty is due to uncertainty
on the theoretical normalization of the background rates. Finally, we apply the unified ordering principle~\cite{feldcous} for the Neyman construction to create confidence intervals in the true value of $\sigma_{N2}$ for each $N_2,N_1$ mass point.

\begin{figure}[!h]
\begin{center}
\includegraphics[width=0.95\linewidth]{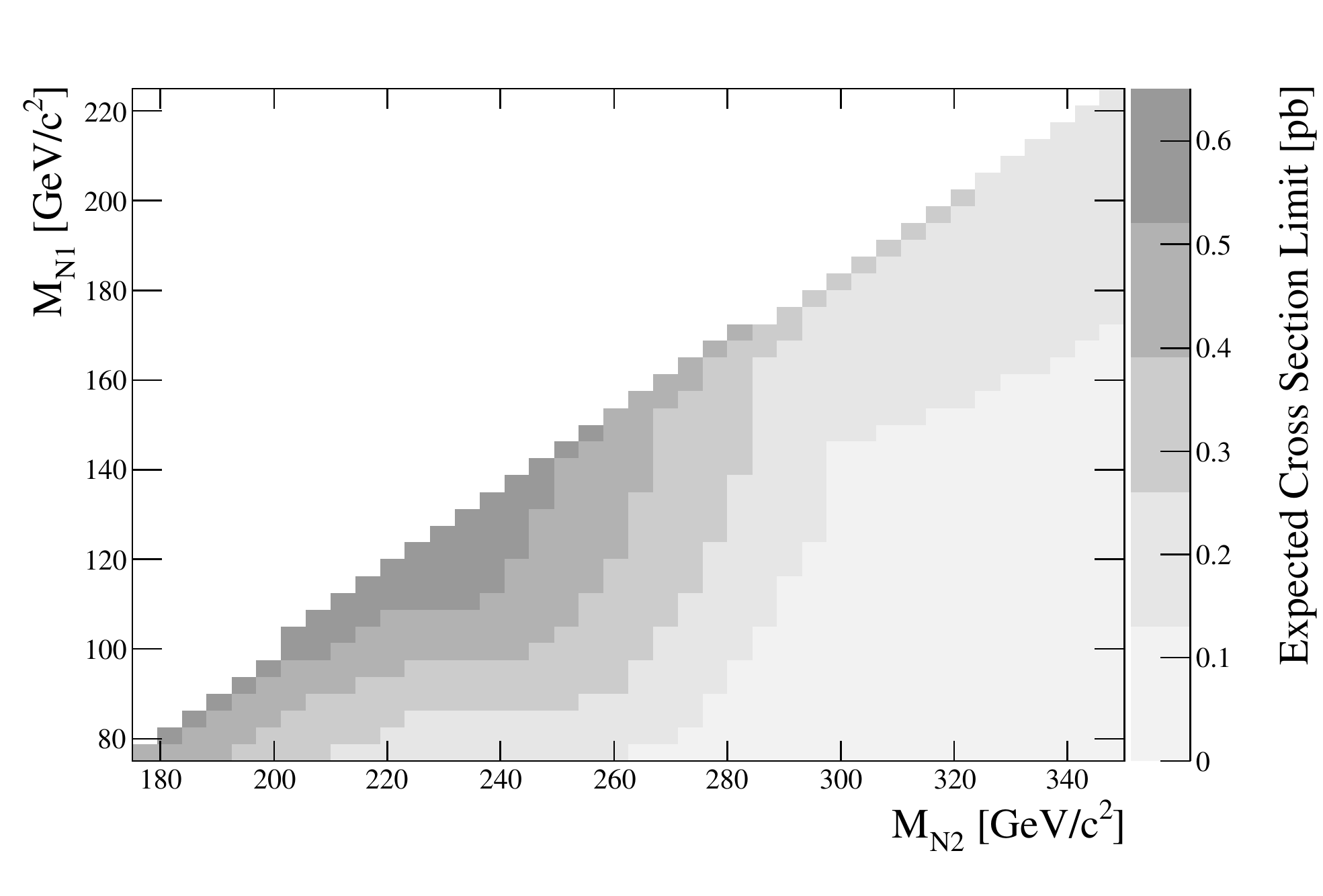}
\includegraphics[width=0.95\linewidth]{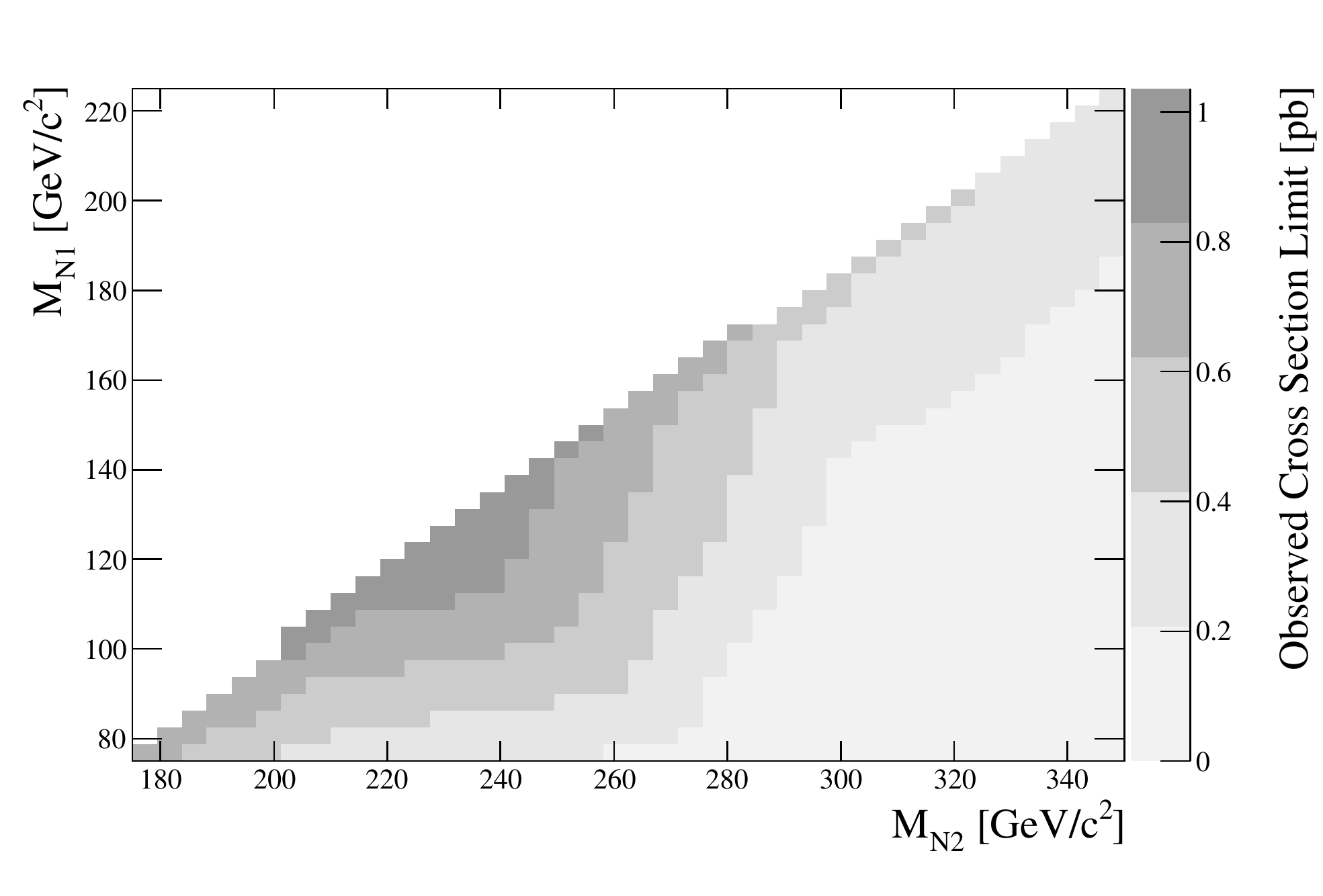}
\caption{ Upper limit at 95\% CL on the cross section of $p\bar{p} \rightarrow
 N_2N_2\rightarrow N_1ZN_1Z$ in data with 4.2 fb$^{-1}$ of integrated
 luminosity as a function of the masses of $N_1$ and
 $N_2$. Top shows median expected limits; bottom shows observed
 limits; see Table~\ref{tab:limits}.}
\label{fig:lims}
\end{center}
\end{figure}

We find the candidate events in the data to be consistent with
expected standard-model backgrounds and thus set upper limits at 95\%
CL on the cross section for $p\bar{p}\to N_2 N_2\to N_1 Z N_1 Z$. Theoretical cross sections for each mass point are presented in Table~\ref{tab:limits}, along with their respective expected and observed limits in our data sample. The expected and observed cross section limits can be seen in Figure~\ref{fig:lims} and Table~\ref{tab:limits}.

\begin{table}[h!]\footnotesize
\caption{ Acceptance of the $ZZ+\missET$ selection for varying
 thresholds in $\missET$ optimized for each point in the $M_{N1}$,
 $M_{N2}$ mass plane. Also shown are the median expected and
 observed 95\% CL upper limits on the cross section ($\sigma_{N2}$)  in data with 4.2 fb$^{-1}$ of integrated luminosity, as well as the
 theoretical prediction~\cite{n2n1theory,madgraph}. }
\begin{center}
\setlength{\tabcolsep}{3pt}
\begin{ruledtabular}
\begin{tabular}{ c c c c c }
$M_{N1}$, $M_{N2}$ & $\missET$ Cut & Acceptance & \multicolumn{2}{c}{ $\sigma_{N2}$ [fb]} \\
\scriptsize [GeV$/c^{2}$] & \scriptsize [GeV] & \scriptsize[\%] & \scriptsize Theory & \scriptsize Exp. /Obs. Limit \\
\hline
75, 175 & 37 & 0.99 & 0.51 & 511 / 702 \\
75, 200 & 68 & 1.02 & 0.21 & 292 / 369 \\
125, 225 & 36 & 0.85 & 0.16 & 684 / 1088 \\
75, 225 & 92 & 0.93 & 0.081 & 156 / 273 \\
75, 275 & 118 & 1.01 & 0.015 & 94 / 132 \\
125, 300 & 119 & 1.06 & 0.013 & 99 / 138 \\
175, 300 & 80 & 0.96 & 0.022 & 171 / 315 \\
125, 350 & 156 & 1.05 & 0.003 & 75 / 48 \\
225, 350 & 80 & 1.05 & 0.006 & 190 / 297 \\
75, 350 & 167 & 1.06 & 0.001 & 71 / 55 \\
\end{tabular}
\end{ruledtabular}
\end{center}
\label{tab:limits}
\end{table}

In summary, we have performed the first search for new phenomena in events with two reconstructed $Z$ bosons and large missing transverse momentum. This signature is sensitive to
processes $p\bar{p}\rightarrow X_2X_2 \rightarrow Z Z X_1 X_1$, where $X_2$ is
an unstable particle decaying as $X_2\rightarrow ZX_1$ and $X_1$ being undetected. The particles $X_1$ and $X_2$
may be, among other possibilities, fourth generation neutrinos or
supersymmetric particles. A specific model in which $X_2$ and $X_1$
are fourth-generation neutrinos is used without loss of generality.  In the final state in which one $Z$ boson decays to
two charged leptons and the second decays hadronically, we find
agreement between the data and the standard-model expectation using
data from proton-antiproton collisions with 4.2 fb$^{-1}$ of integrated luminosity. 
Based on the results in Table~\ref{tab:limits}, we report 95\% CL upper limits on the cross section of the process
$p\bar{p}\rightarrow X_2X_2 \rightarrow Z Z X_1 X_1$ ranging from 50 fb to 1
pb depending on the masses of $X_1$ and $X_2$.

\newpage

We thank the Fermilab staff and the technical staffs of the
participating institutions for their vital contributions. This work
was supported by the U.S. Department of Energy and National Science
Foundation; the Italian Istituto Nazionale di Fisica Nucleare; the
Ministry of Education, Culture, Sports, Science and Technology of
Japan; the Natural Sciences and Engineering Research Council of
Canada; the National Science Council of the Republic of China; the
Swiss National Science Foundation; the A.P. Sloan Foundation; the
Bundesministerium f\"ur Bildung und Forschung, Germany; the Korean World Class University Program, the National Research Foundation of Korea; the Science and Technology Facilities Council and the Royal Society, UK; the Institut National de Physique Nucleaire et Physique des Particules/CNRS; the Russian Foundation for Basic Research; the Ministerio de Ciencia e Innovaci\'{o}n, and Programa Consolider-Ingenio 2010, Spain; the Slovak R\&D Agency; and the Academy of Finland.


\begin{thebibliography}{36}
\bibitem{coordinates}
CDF uses a cylindrical coordinate system with
 the $z$ axis along the proton beam axis. Pseudorapidity is $\eta
 \equiv -\ln(\tan (\theta/2))$, where $\theta$ is the polar angle
 relative to the proton beam direction, and $\phi$ is the azimuthal angle while $p_T = |p|\ \sin\theta$, $E_T = E\ \sin\theta$.
\bibitem{metdef} 
Missing transverse momentum, $\missET$, is defined as the vector $ -\sum_i E_T^i \vec n_i$ where $E_T^i$ are the magnitudes of transverse momentum contained in each calorimeter tower $i$, and $\vec n_i$ is the unit vector from the interaction vertex to the tower in the transverse ($x,y$) plane.
\end{thebibliography}
\end{document}